\documentclass[reprint,superscriptaddress,amsmath,amssymb,aps,]{revtex4-2}
\usepackage{graphicx}
\usepackage{graphicx}
\usepackage{dcolumn}
\usepackage{bm}
\usepackage{physics}
\usepackage{xcolor}
\definecolor{darkblue}{rgb}{0,0,0.6}
\usepackage[colorlinks,linkcolor=darkblue,citecolor=darkblue,urlcolor=darkblue]{hyperref}
\usepackage[colorlinks,linkcolor=darkblue,citecolor=darkblue,urlcolor=darkblue]{hyperref}

\newcommand{\eps}{\varepsilon}
\newcommand{\Ti}{T_i} 
\newcommand{\Ta}{T_a} 
\newcommand{\Tg}{T_g} 
\newcommand{\ttransf}{t_\mathrm{tr}} 
\newcommand{\tB}{\tau_B} 
\newcommand{\talpha}{\tau_\alpha} 
\newcommand{\CB}{C_B} 
\newcommand{\xl}{x_l}       
\newcommand{\lphtr}{\ell_\mathrm{tr}} 
\newcommand{\lcross}{\ell_c}

\newcommand{\montpellier}{Laboratoire Charles Coulomb (L2C), Universit\'e de Montpellier, CNRS, 34095 Montpellier, France}
\newcommand{\cambrigeM}{Department of Applied Mathematics and Theoretical Physics,
University of Cambridge, Wilberforce Road, Cambridge CB3 0WA, United Kingdom}
\newcommand{\cambridgeQ}{Yusuf Hamied Department of Chemistry, University of Cambridge,
Lensfield Road, Cambridge CB2 1EW, United Kingdom}
\newcommand{\madison}{Department of Chemistry, University of Wisconsin–Madison, Madison, Wisconsin 53706, USA}

\begin{document}

\title{Two-step devitrification of ultrastable glasses}

\author{Cecilia Herrero}

\affiliation{\montpellier}

\author{Camille Scalliet}

\affiliation{\cambrigeM}

\author{M. D. Ediger}

\affiliation{\madison}

\author{Ludovic Berthier}

\affiliation{\montpellier}

\affiliation{\cambridgeQ}

\date{\today}

\begin{abstract}
  The discovery of ultrastable glasses has raised novel challenges about glassy systems. Recent experiments studied the macroscopic devitrification of ultrastable glasses into liquids upon heating but lacked microscopic resolution. We use molecular dynamics simulations to analyse the kinetics of this transformation. In the most stable systems, devitrification occurs after a very large time, but the liquid emerges in two steps. At short times, we observe the rare nucleation and slow growth of isolated droplets containing a liquid maintained under pressure by the rigidity of the surrounding glass. At large times, pressure is released after the droplets coalesce into large domains, which accelerates devitrification. This two-step process produces pronounced deviations from the classical Avrami kinetics and explains the emergence of a giant lengthscale characterising the devitrification of bulk ultrastable glasses. Our study elucidates the nonequilibrium kinetics of glasses following a large temperature jump, which differs from both equilibrium relaxation and aging dynamics, and will guide future experimental studies.
\end{abstract}

\maketitle

Glasses play a critical role in modern technology, in applications as diverse as optical fibers and organic light emitting diode (OLED) displays~\cite{berthier2016facets,rafols2018high}. However, in spite of their solid-like appearance, all glasses spontaneously evolve in a process known as physical aging~\cite{angell2000relaxation}. Aging is an inherent feature of the glassy state that results from its nonequilibrium nature, and it is an important challenge because the resulting structural evolution can negatively influence material properties. As a result of recent progress, aging is reasonably well understood if the temperature is lowered by only a few Kelvins near equilibrium. In this case, knowledge of equilibrium response functions allows predictions for the aging dynamics~\cite{riechers2022predicting}. Larger jumps to low temperature are more challenging, but their qualitative features are well understood~\cite{Scherer1990,Hodge1994,angell2000relaxation,simon2002enthalpy,riechers2022predicting}.

By contrast, the opposite case of a large positive temperature jump is much less understood, and enhanced understanding could provide important conceptual advances regarding amorphous materials.  Qualitatively different phenomena are observed in comparison to the down jumps and near-equilibrium jumps, with structural relaxation at first occurring very slowly and then accelerating~\cite{angell2000relaxation,kovacs1964transition}. The recent development of ultrastable glasses, accessed via new experimental~\cite{Swallen2007,ediger2017perspective,rodriguez2022ultrastable} and simulation~\cite{berthier2016equilibrium,Ninarello2017,parmar2020ultrastable} techniques, provides a new tool to understand large up-jump experiments. Because these materials are prepared in low energy states with high kinetic barriers, extreme up-jump experiments can now be performed with a wider variety of experimental techniques.

In the last decade, positive temperature jumps performed from ultrastable states have revealed previously hidden features of amorphous materials. For ultrastable films, the devitrification from the glass to the liquid occurs heterogeneously, starting from the free surface and propagating towards the interior over large distances ($>$ 1 micron). This process has been studied extensively in experimental~\cite{Kearns2010,walters2014thermal,Sepulveda2014,rafols2017role,RodriguezTinoco2019} and theoretical~\cite{Wolynes2009,leonard2010macroscopic,Wisitsorasak2014,Gutierrez2016,Flenner2019} work.

\begin{figure*}[t]
\centering
\includegraphics[width=\textwidth]{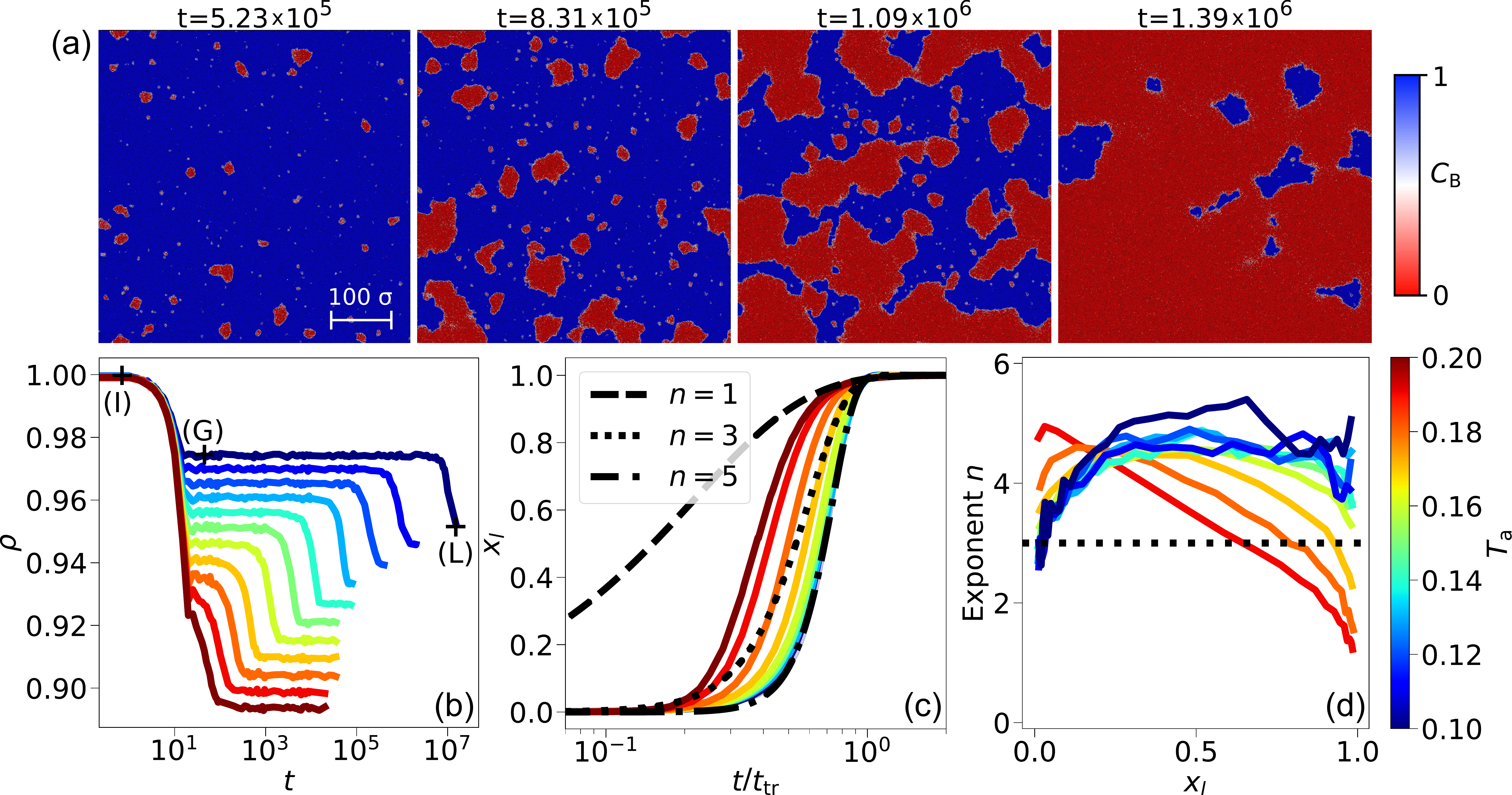}
\caption{{\bf Transformation of an ultrastable glass into an equilibrium liquid.}
  (a) Visualising the transformation of an ultrastable glass prepared at $T_i=0.035$ annealed at $\Ta=0.11$ with $N=256000$ at times $t$ indicated on the figure. The bond-breaking correlation distinguishes the glass with $C_B^i=1$ (blue) from the liquid with $C_B^i=0$ (red). Rare isolated liquid droplets nucleate, grow slowly, then coalesce to transform the entire system.    
  (b) Time evolution of number density $\rho$ for $T_i=0.035$ and different annealing temperatures $\Ta$. The initial glass (I) rapidly expands to the glass state (G), and transforms over a much slower time scale into the equilibrium liquid (L).  
  (c) Time evolution of the liquid fraction $x_l(t)$ for the same parameters as (b). The time axis is normalised by the transformation time $\ttransf$. Dashed lines: Avrami functions $1-\exp(-K t^n)$ with different exponents $n$.
  (d) Evolution of the effective exponent $n$ as a function of $\xl(t)$. The dashed classical Avrami value $n=3$ does not adequately describe our results.}
\label{fig:fig1}
\end{figure*}

Fewer studies have investigated the homogeneous devitrification of ultrastable glasses in the bulk~\cite{sepulveda2013manipulating,rodriguez2016relaxation,Rafols2018,VilaCosta2020}. Surprisingly, macroscopic measurements are best interpreted using an analogy with the nucleation and growth kinetics traditionally observed across first-order phase transitions~\cite{Kearns2010,Jack2016}. However, quantitative analysis requires a number of assumptions~\cite{Kearns2010,VilaCosta2020} regarding the underlying physics, which have not been tested. Previous simulations used relatively small systems~\cite{Hocky2014equilibrium,Fullerton2017} or simplified models~\cite{Jack2016,Gutierrez2016,Lulli2020spatial}. Therefore, an understanding of microscopic processes, of the characteristic time scales and length scales and their evolution with glass stability is lacking. At the fundamental level, confirming a deep analogy between crystal melting and devitrification would provide experimental evidence of a thermodynamic interpretation of the glass transition~\cite{Jack2016,berthier2016facets}.

We report results from molecular dynamics simulations which paint a complete microscopic description of the transformation kinetics of ultrastable glasses following a positive temperature jump. We prepare ultrastable configurations at very low initial temperature $\Ti$ using the swap Monte Carlo algorithm~\cite{Ninarello2017,Berthier2019swap}. We instantaneously heat the system to the annealing temperature $\Ta > \Ti$ and observe how the system devitrifies from its initial low-enthalphy state to the equilibrium liquid at $\Ta$. Our results generally confirm a nucleation and growth kinetics, but they deviate from the classical Avrami description~\cite{Avrami1939} and existing interpretations on several aspects. In particular, the liquid droplets that nucleate initially grow much more slowly than the large domains found at larger times, because they contain a liquid that is compressed by the surrounding glass. As a result, the large length scale characterising the bulk devitrification of ultrastable glasses is a novel physical quantity, distinct from the crossover length scale discussed for thin films. 

We perform molecular dynamics (MD) simulations of a two dimensional size-polydisperse mixture of $N=64000$ soft repulsive particles, known to be a reliable glass-former~\cite{Berthier2019poly2d}. We prepare ultrastable configurations at number density $\rho_{i} = 1$, temperature $\Ti = 0.035$ and pressure $P_i=2.31$ using the swap Monte Carlo algorithm~\cite{Ninarello2017,Berthier2019swap} and reduced numerical units. We then perform classical MD simulations in the $NPT$ ensemble at pressure $P_i$ for a broad range of annealing temperatures $\Ta > \Ti$. More information about the model, simulations and reduced units are provided in the Methods. 

We start in Fig.~\ref{fig:fig1}a with a global description of the transformation process for the particular case $\Ta=0.11$. To visualise the transformed regions, we introduce a local structural correlation function $C_B^i(t)$ which records the fraction of nearest neighbors lost by particle $i$ since $t=0$~\cite{scalliet2022thirty}. Below, we count a particle as `liquid' whenever $C_B^i(t) \leq 0.5$, but we checked equivalence with several other definitions. We observe that devitrification starts from rare regions distributed throughout the system. As time increases, the size of these regions grows and new regions keep appearing. This is consistent with nucleation and growth, but distinct from the pre-existing nuclei picture proposed recently~\cite{VilaCosta2020}. At larger times, the growing droplets coalesce and then percolate throughout the system. At very late times, a few glass regions survive, which are eventually invaded by the surrounding liquid. We define the characteristic times $\ttransf$ and $t_{1/2}$ when respectively $99\%$ and $50\%$ of the particles have become liquid. 

The time evolution of the density is shown in Fig.~\ref{fig:fig1}b for several annealing temperatures $\Ta$. By construction, all curves start from the initial density $\rho_i=1$, state (I). As a result of the temperature jump from $\Ti$ to $\Ta$, the system quickly expands without structural rearrangement to reach the glass state (G) in Fig.~\ref{fig:fig1}b. Its density is smaller, $\rho_G < \rho_i$. At much larger times, the glass transforms into the final liquid (L) which has an even lower density $\rho_L < \rho_G$. Both $\rho_G$ and $\rho_L$ depend on $\Ta$. As found in smaller temperature jumps~\cite{angell2000relaxation,kovacs1964transition}, the dynamics is initially very slow but the glass transformation suddenly accelerates and takes place abruptly. The transformation time $\ttransf$ increases by 5 orders of magnitude with decreasing the annealing temperature $\Ta$.

\begin{figure*}[t]
 \centering
 \includegraphics[width=\textwidth]{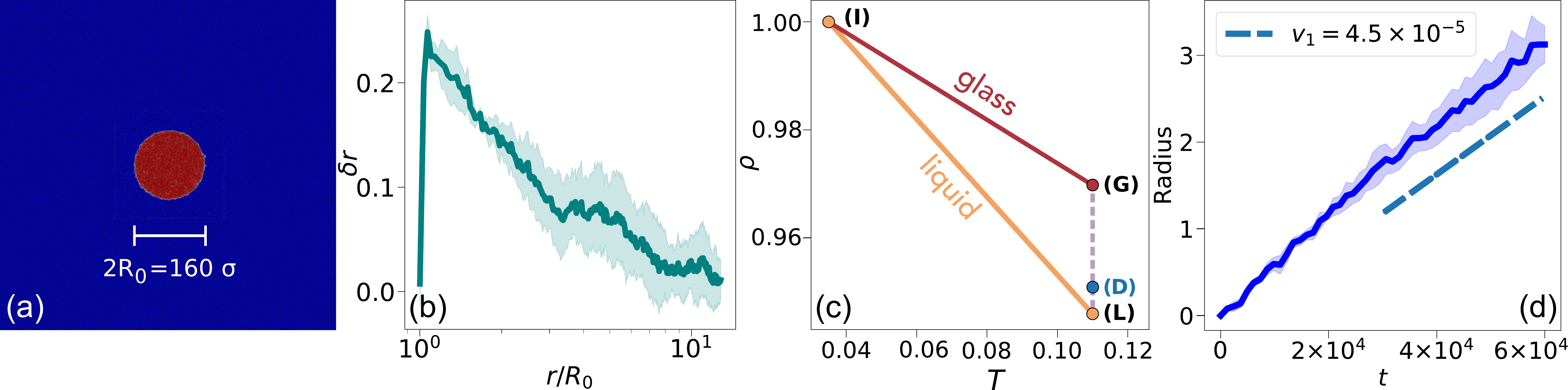}
 \caption{{\bf Understanding the early stages of the transformation by simulating isolated droplets.}
   (a) Snapshot of a single liquid droplet of radius $R_0=80$ in a large $N=576000$ sample. Color code provided in Fig.~\ref{fig:fig1}a. 
   (b) Radial displacement field in the glass after mechanical equilibrium is reached for $R_0=30$. 
   (c) Liquid and glass equations of state showing initial state (I), the expanded glass (G) and final liquid (L) for $T_i=0.035$ and $T_a=0.11$. (D) indicates the isolated droplet density. 
   (d) The time evolution of the droplet radius allows us to measure the growth velocity $v_1$ of isolated droplets containing the compressed liquid.}
\label{fig:fig2}
\end{figure*}

The transformation kinetics can be quantified using the fraction $x_l(t)$ of liquid particles at time $t$, which goes from 0 at $t=0$ to 1 at long times, see Fig.~\ref{fig:fig1}c. To compare different $\Ta$, we rescale the time axis by $\ttransf$. As $\Ta$ decreases, the time evolution of $\xl(t)$ becomes sharper and eventually collapses onto a master curve for the lowest annealing temperatures. 

Following previous studies~\cite{Kearns2010,Gutierrez2016,VilaCosta2020}, we fit $x_l(t)$ to the Avrami equation~\cite{Avrami1939} 
\begin{equation}
    \xl(t) = 1 - \exp(-K t^n).
    \label{eq:Avrami}
\end{equation}
In the classical Avrami picture, a nucleation rate describes the initiation of droplets, and a constant velocity captures their growth. Both parameters enter the definition of $K$, and the exponent is predicted to be $n=d+1$, with $d$ the spatial dimension. The classical Avrami prediction is thus $n=3$ for our two-dimensional simulations. We show in  Fig.~\ref{fig:fig1}c that this exponent does not describe our measurements accurately. To better quantify these deviations, we follow the experimental literature and report the parametric evolution of the effective exponent $n$ as a function of $x_l$ in Fig.~\ref{fig:fig1}d. While $n$ is not well-defined at high $T_a$, it consistently takes a surprisingly large value $n \approx 4.5$ for low $T_a$, indicating a breakdown of the classical nucleation and growth picture. In the following we deploy original numerical strategies to explain this behaviour.

In the early stages of the transformation, the system is composed of isolated liquid droplets confined by the surrounding glass matrix. We analyse an idealised version of this geometry with a single liquid droplet of radius $R_0 \gg 1$ immersed in a very large glass matrix of linear size $L \gg R_0$, see Fig.~\ref{fig:fig2}a. We used $L\approx 770$ and found identical results for $R_0=30$, 80, 50 and 100. Our protocol, detailed in the Methods, is the following. We start from the glassy state (G) in Fig.~\ref{fig:fig1}b and perform high-temperature dynamics in a spherical domain of radius $R_0$ to transform this region, followed by thermalization at $\Ta$. This is done while keeping the particles outside the cavity immobile. At this stage, the density in the liquid droplet is thus equal to that of the glass (G), implying that the liquid pressure has increased above $P_i$. Then, we simulate the dynamics of the whole sample at temperature $T_a$ and constant volume. At the very beginning of the simulation, the compressed liquid exerts pressure on the surrounding glassy particles to expand, but the expansion is opposed by the rigidity of the glass, until a mechanical equilibrium is found.  We have computed the radial displacement in the glass right after mechanical equilibrium is reached. Its angular-averaged amplitude $\delta r$, shown in Fig.~\ref{fig:fig2}b, is maximal at the edge of the droplet $r/R_0 = 1$ and decays algebraically to zero at large distances. Consequently, the droplet expands from $R_0$ to $R_0 + \Delta R_0$ so that its density decreases by $\Delta \rho \approx 2 \rho_G \Delta R / R_0$. The corresponding droplet (D) density $\rho_D = \rho_G - \Delta \rho$ is shown in Fig.~\ref{fig:fig2}c together with the other state points. Far from the droplet, the glass pressure is equal to $P_i$, but it is larger in the liquid droplet which contains a liquid at density $\rho_D$ and pressure $P_D > P_i$ which is therefore distinct from the final equilibrium liquid (L) which is instead at conditions $(P_i, T_a)$.

At much longer times, structural relaxation occurs and the liquid droplet grows slowly. We monitor the time evolution of the liquid fraction $x_l(t)$, which is directly connected to the growing radius $R(t)$ since $\xl(t) \propto R^2(t)$. In Fig.~\ref{fig:fig2}d we show that the droplet size increases linearly with time, $R(t) = R_0 + v_1 t$, with a velocity $v_1$ that can be measured numerically. We have confirmed that the same value $v_1$ describes the growth of isolated droplets randomly selected in the real process illustrated in Fig.~\ref{fig:fig1}. This demonstrates that the idealised geometry depicted in Fig.~\ref{fig:fig2} faithfully represents the initial stages of the bulk devitrification.

\begin{figure}[t]
 \centering
 \includegraphics[width=\linewidth]{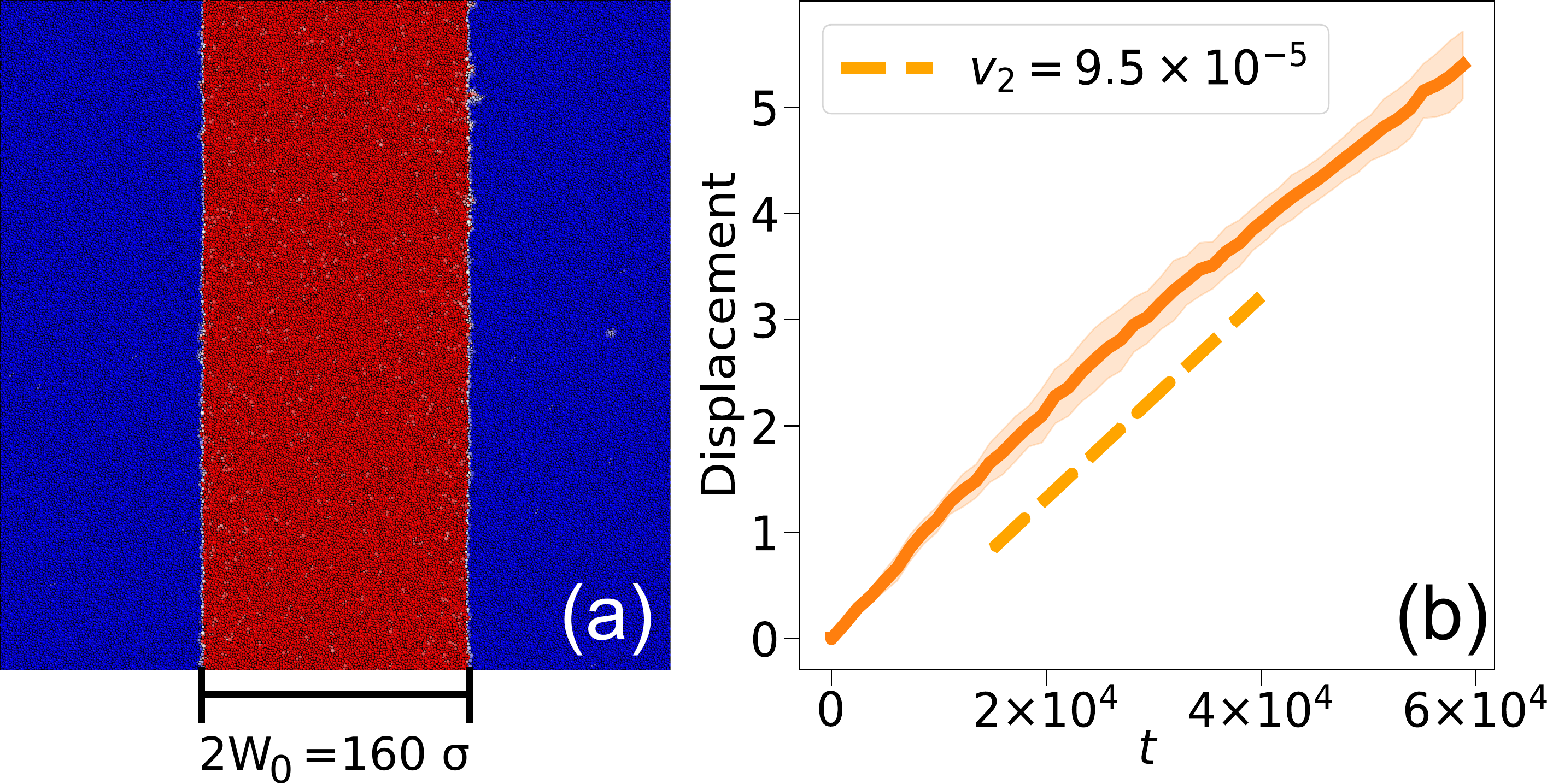}
 \caption{{\bf Understanding the late stages of the transformation by simulating a macroscopic interface.}
   (a): Snapshot of a two-front system with $N=64000$. Color code as in Fig.~\ref{fig:fig1}a.
   (b) The time evolution of the interface position allows us to measure the growth velocity $v_2$ of large fronts.}
 \label{fig:fig3}
\end{figure}

To better characterize the late times at which very large liquid domains percolate throughout the system, we simulate the idealised geometry shown in Fig.~\ref{fig:fig3}a in which we create a single percolating liquid domain~\cite{Hocky2014equilibrium} of width $2 W_0$. The numerical strategy is similar to that for the droplet geometry, see Methods. Once the vertical liquid domain is prepared, we run $NPT$ simulations at temperature $\Ta$ and pressure $P_i$ for the whole sample. In this case, the pressure is homogeneous across the system and the liquid density is $\rho_L$. As time increases, we observe that the liquid propagates towards the glass as a front. We extract its velocity by monitoring the time evolution of $x_l(t)$, which is directly proportional the displacement of the interface. This is shown in Fig.~\ref{fig:fig3}b for $\Ta = 0.11$. The front propagation is again ballistic, with $x_l(t) \propto v_2 t$ which defines the front velocity $v_2$. Again, we have checked that the same value $v_2$ describes the growth of large interfaces randomly selected in the real transformation process shown in Fig.~\ref{fig:fig1}, confirming the relevance of the idealised geometry studied in Fig.~\ref{fig:fig3}. 

By gathering velocity measurements for different annealing temperatures and the two idealised geometries shown in Figs.~\ref{fig:fig2}, \ref{fig:fig3} we find that the growth of large interfaces is systematically faster than the growth of isolated droplets, $v_1 < v_2$. This is because the liquid inside droplets is under pressure $P_D > P_i$ and its relaxation time is thus larger than the one of the equilibrium liquid (L). For a given $\Ti$, we observe that the difference $v_2-v_1$ increases with increasing $\Ta$, because the separation between the points (G) and (L) increases in Fig.~\ref{fig:fig2}c. For a given $\Ta$, we also find a larger difference in the velocities for lower $\Ti$ because the glass becomes stiffer and compresses the liquid droplet more at short times. 

The quantitative comparison between the kinetics of isolated droplets and large fronts demonstrates that a central assumption leading to Avrami kinetics does not hold for the bulk devitrification of ultrastable glasses, as there is not a unique velocity describing the entire process. We argue that this more complex two-step process is responsible for the anomalously large value of the exponent $n$ shown in Fig.~\ref{fig:fig1}e. First, we note that at early times $x_l \ll 1$ when only isolated droplets exist, the classical value $n = 3$ is observed. However, as the liquid fraction grows and droplets start to coalesce, $n$ grows to a much larger value $n \approx 4.5$. Qualitatively, this implies that the transformation process accelerates further, as expected if the velocity increases from $v_1$ to $v_2$. In the Supplementary Information, we show that adding the simplest hypothesis of a smooth transition between two values $v_1$ and $v_2>v_1$ in the analytic Avrami description generically produces values $n \approx 5$, which supports our physical interpretation of the numerical results. Note that the pressure effect, which controls kinetic stability and the devitrification melting, is absent in several simulation studies~\cite{Hocky2014equilibrium,Jack2016,Gutierrez2016,Lulli2020spatial}.   
\begin{figure*}[t]
\centering
\includegraphics[width=\textwidth]{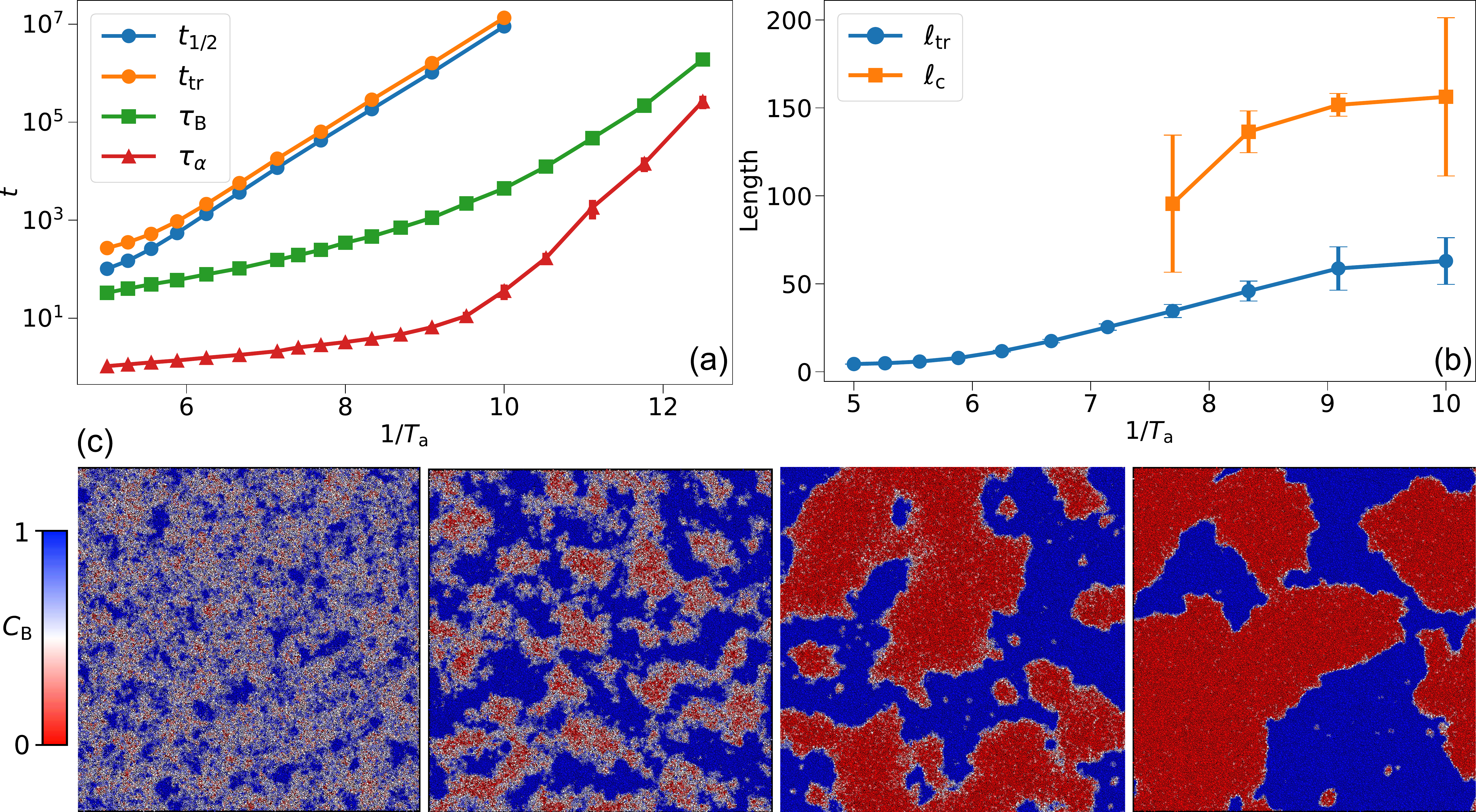}
\caption{{\bf Characteristic time scales and length scales of ultrastable glass transformation.}
  (a) Temperature dependence of the transformation time $\ttransf$ and two characteristic times of the equilibrium liquid: $\tB$ defined from $\CB$ and $\talpha$ from $F_s(q,t)$.
  (b) Characteristic lengths of the process: $\lphtr$ is the characteristic size of the domains at time $t_{1/2}$, and the crossover length $\lcross = v_2 \ttransf$.
  (c) Snapshots taken at time $t_{1/2}$ for $N=64000$ and different annealing temperatures $\Ta = 0.20$, 0.17, 0.13, 0.10 (from left to right).}
\label{fig:fig4}
\end{figure*}

Finally, we collect in Fig.~\ref{fig:fig4} the relevant time scales and length scales characterising the devitrification of ultrastable glasses. Starting with times in Fig.~\ref{fig:fig4}a, the slowest time scale is the transformation time $\ttransf$, which is much larger than any time scale characterising the equilibrium relaxation of the liquid in the same conditions. We display the decay time of the bond-breaking correlation $C_B(t)$, $\tau_B$, and of the self-intermediate scattering function, $\tau_\alpha$, defined in the Methods section. The kinetic stability, encoded in the ratio between $\ttransf$ and these equilibrium relaxations reaches a large value $10^3-10^5$ at the lowest $T_a$, and would increase further if longer timescales could be simulated allowing access to even lower $\Ta$. This large kinetic stability is comparable to values previously reported in both vapor deposited films~\cite{Dawson2012,Sepulveda2014,RodriguezTinoco2019} and simulations performed using the swap Monte Carlo algorithm~\cite{Fullerton2017,Flenner2019}.

Turning to length scales, we report in Fig.~\ref{fig:fig4}b the evolution of the characteristic transformation length $\lphtr$ suggested by Fig.~\ref{fig:fig4}c where snapshots of the system at various $T_a$ are shown at time $t_{1/2}$ when half of the glass has transformed. These images reveal that the transformation process is highly heterogeneous in space, with a characteristic length scale $\lphtr$ which grows as kinetic stability increases. To characterize this length, we employ the method using chord length distributions, following the same recipe as in \cite{Testard2011,Testard2014}. We define $\lphtr$ as the first moment of the chord distributions. As shown in Fig.~\ref{fig:fig4}b, $\lphtr$ grows with decreasing $\Ta$ reaching $\lphtr \approx 60$ for $\Ta=0.10$ and a kinetic stability of about $4 \times 10^5$. In these conditions, the transformed domains barely fit in our simulation box of linear size $L\approx 250$, see the rightmost snapshot in Fig.~\ref{fig:fig4}c.

In experimental studies of thin films transforming via front propagation, a large `crossover' length scale $\ell_c$ has been reported~\cite{Kearns2010,RodriguezTinoco2019}. It is defined as the distance travelled by the liquid front over the transformation time of the corresponding bulk system. Using our notations, we have $\lcross = v_2 \ttransf$, which can be determined from our independent measurements of $v_2$ and $\ttransf$. Previous work~\cite{Kearns2010,RodriguezTinoco2019,VilaCosta2020} assumed that the two length scales $\lphtr$ and $\lcross$ are equal. This is indeed correct when the classical Avrami picture applies~\cite{Jack2016}. Our results however imply that this picture does not describe the transformation of ultrastable glasses. In Fig.~\ref{fig:fig4}b we confirm that the two characteristic length scales are different, and evolve differently with $\Ta$. Again, the existence of two distinct velocities rationalises this finding. A smaller value $v_1 < v_2$ in the first stages of devitrification implies that $\ttransf$, and thus $\lcross$, are larger than if $v_1 = v_2$. Instead, a longer $\ttransf$ allows more time for droplets to nucleate so that $\lphtr$, which controlled by the typical distance between droplets, decreases. The opposite influence of the velocity contrast $v_2/v_1$ on physical length scales implies that in general $\lphtr$ is distinct from, and smaller than, $\ell_c$. 

Our computational study of the devitrification of ultrastable glasses reveals several unexpected results that had not been predicted by previous theoretical models, and have either not been observed experimentally or contradict published results. We now discuss these points.

Our first novel observation is the nucleation at short times of liquid droplets containing a liquid that is under pressure. This stems from the large density difference between the ultrastable glass and the equilibrium liquid, combined to the rigidity of the surrounding glass, see Fig.~\ref{fig:fig2}. Although these two effects may quantitatively differ in different materials, it is difficult to imagine a physical relaxation process that would entirely release this pressure. This finding has several important consequences. First, it implies that the liquid that first appears during the transformation has a relaxation dynamics that is much slower than the equilibrium liquid that forms at long times. Published works probing these dynamics have not reported any evidence for such a slow process~\cite{Sepulveda2014,Rafols2018,VilaCosta2020}, and future experiments should revisit this point. A related consequence is that the growth of these compressed liquid droplets is much slower than the growth of the fronts propagating from a free surface. This directly contradicts an important  assumption made in Ref.~\cite{Kearns2010,VilaCosta2020} to interpret experimental results. To our knowledge, the growth of isolated droplets has not yet been studied experimentally. 

A final consequence is that the growth velocity of the liquid phase is not constant during the transformation, which directly explains the emergence of an anomalously large Avrami exponent describing the transformation kinetics. This finding constitutes our second key novel observation. It contrasts with the recent report of a smaller exponent extracted from calorimetric measurements~\cite{VilaCosta2020}. These recent results were explained by hypothesizing a population of pre-existing defects initiating the transformation. We have not observed such defects in our simulations where the appearance of new liquid droplets is instead slow and randomly distributed in time. We can only invoke two important differences between simulations and experiments that could explain this discrepancy. We directly determine the liquid fraction $x_l(t)$ {\it in situ} whereas in experiments it is indirectly inferred from calorimetric measurements involving a complicated thermal treatment and data analysis. Another difference is the devitrification time window analysed in simulations, which is of order 10~ms in simulations, but about $10^3$ times slower in most experiments~\cite{VilaCosta2020}.

Our third key observation is that a large length scale characterizes the transformation process, which becomes larger as the kinetic stability of the glass increases. Despite our numerical constraints, we directly measure a length scale approaching $\ell_{\rm tr} \approx 100 \sigma$ which would presumably grow larger if we could simulate longer time scales. We are not aware of any bulk relaxation process in equilibrium supercooled liquids or in aging glasses that can reach such large values. Qualitatively, the large length scale $\lphtr$ reflects the equally-large distance between nucleated liquid droplets in the early stages of the transformation. As $T_a$ decreases, droplets become more sparse as their nucleation rate decreases, and $\lphtr$ increases. Therefore, it is the large barrier to nucleating liquid droplets which is responsible for the emergence of a large length scale. This interpretation explains why such large length scales have never been found in conventional glasses, as these less stable systems would have smaller barriers.  

There has been extensive experimental work deducing that a large crossover length scale $\ell_c$ appears in thin films transformation via the propagation of a front. We have shown that these two length scales are conceptually and quantitatively distinct. In particular, $\ell_c$ simply represents a crossover between distinct physical regimes during front propagation, but does not characterise any kind of spatial correlations, contrary to $\ell_{\rm tr}$. The length scale $\ell_{\rm tr}$ has recently been indirectly inferred from experiments, but this assumes hypotheses which do not hold in our simulations. A more direct experimental measurement of the large length scale $\ell_{\rm tr}$ is an important target for future work.

More fundamentally, the observation that the transformation of an ultrastable glass into a liquid proceeds via a physical mechanism traditionally observed for first-order transitions raises important questions. It was argued in Ref.~\cite{Jack2016} that the existence of a thermodynamic Kauzmann transition provides the needed ingredients to interpret this analogy. In the alternative approach of Ref.~\cite{Gutierrez2016} based on a kinetically constrained model, the additional introduction of some thermally activated nucleation sites was needed to reproduce the observed phenomenology, with the growth processes controlled by dynamic facilitation. Future work should therefore concentrate on the first stages of the transformation to understand better their physical origin. It would also be very interesting to understand whether the regions that first transform are related to specific structural features of ultrastable glassy states. 

\section*{Acknowledgments}

This work was publicly funded through ANR (the French National Research Agency) under the Investissements d'avenir programme with the reference ANR-16-IDEX-0006. It was also supported by a grant from the Simons Foundation (\#454933, LB), the European Research Council under the EU’s Horizon 2020 Program, Grant No. 740269 (CS), the U.S. National Science Foundation, CHE-2153944 (MDE), and by a Visiting Professorship from the Leverhulme Trust (VP1-2019-029, LB). CS acknowledges support from a Herchel Smith Fellowship, University of Cambridge, and a Ramon Jenkins Research Fellowship from Sidney Sussex College, Cambridge.

\bibliography{cecilia.bib}

\newpage

\section*{Methods}

\label{sec2}

\textbf{Size-polydisperse model glass-former.} We perform molecular dynamics (MD) simulations of a two dimensional size-polydisperse mixture, which is well characterised~\cite{Berthier2019poly2d}. The system is composed of soft repulsive spheres whose diameters $\sigma_i$ follow the probability distribution $\mathcal{P}(\sigma_i) = A \sigma_i^{-3}$, with $A$ a normalization constant, $\sigma_i \in [ \sigma_\mathrm{min}, \sigma_\mathrm{max} ]$, $\sigma_\mathrm{min}/\sigma_\mathrm{max} = 0.45$, and $\sigma_\mathrm{max}=1.62$. The pair interaction potential is given by
\begin{equation}
    V_{ij} (r) = \eps \qty(\frac{\sigma_{ij}}{r})^{12} + c_0 + c_2 \qty(\frac{r}{\sigma_{ij}})^2 + c_4 \qty(\frac{r}{\sigma_{ij}})^4;
\end{equation}
where $r = \abs{\vb{r}_i - \vb{r}_j}$ ($\vb{r}_i$ being the position of particle $i$), and non-additive interactions  $\sigma_{ij}=0.5(\sigma_i + \sigma_j)\qty(1 - 0.2 \abs{\sigma_i - \sigma_j})$. We use reduced units based on the particle mass $m$, the energy scale $\eps$, and a microscopic length $\sigma$ defined as the average particle diameter. The time unit is $\tau_\mathrm{LJ} = \sigma \sqrt{m/\eps}$. The parameters $c_0 = -28\eps/r_\mathrm{c}^{12}$, $c_2 = 48\eps/r_\mathrm{c}^{14}$, and  $c_4 = -21 \eps/r_\mathrm{c}^{16}$ are introduced in order to ensure that the potential $V_{ij}$ is continuous up to its second derivative at the cutoff distance $r_\mathrm{c}=1.25~\sigma_{ij}$.

\textbf{Simulating the transformation of an ultrastable glass into a liquid.} The initial configuration is composed of $N=64000$ particles at a number density $\rho_i = N/L^2 = 1$, in a square box of linear size $L$ and periodic boundary conditions. The initial state is equilibrated using the swap Monte Carlo algorithm~\cite{Berthier2019swap}, at a temperature $\Ti = 0.035 \approx \Tg/2$, with $\Tg$ the experimental glass transition temperature. The corresponding pressure is $P_i=2.31$. The MD simulations are performed using a Nos\'e-Hoover thermostat at temperatures $\Ta \in [0.1,0.2]$ and a barostat at pressure $P_i$. The discretisation timestep is set to $0.01$. For each $\Ta$ we perform MD simulations starting from 8 initial independent configurations, with 3 different sets of initial velocities, therefore gathering statistics for 24 different runs per $\Ta$. The errorbars were computed from the statistical error within $95~\%$ of confidence level. In order to test for possible finite size effects, we performed equivalent MD simulations for a system composed of $N=256000$ particles, running 3 different configurations for $\Ta = 0.11$, 0.12, 0.13 and obtained equivalent results.

\textbf{Simulations of isolated droplets.} The key technical aspect of these simulations is to allow the system to relax mechanically and possess an inhomogeneous pressure profile. We have found that the following strategy satisfies our needs. We start from the same initial configurations as in the full phase-transformation simulations using an $NPT$ simulation at $T=\Ta$ and $P_i$. The expanded glass (G) is reached after a short time, shown in Fig.~\ref{fig:fig2}a. Immediately after this rapid expansion, we replicate the system three times in both dimensions, resulting in a glass with $N=576000$ particles with pressure $P_i$. Such a large system size is needed to ensure that the mechanical response of the glass matrix is not affected by finite size effects, or, equivalently to ensure that the liquid droplet does not interact mechanically with its replicated images. We insert a liquid droplet in this large glass matrix by performing high-temperature dynamics, at $T=0.8$ during $10^2~\tau_{\rm LJ}$ in a circular cavity of radius $R_0$, keeping all other particles fixed. This is followed by additional simulations inside the cavity at temperature $\Ta$ for $10^4~\tau_{\rm LJ}$ to ensure equilibration inside the cavity. We then perform $NVT$ simulations of the entire system at $\Ta$. In the first stages of these simulations, the liquid droplet expands into the glass matrix, which pushes back until mechanical equilibrium is reached, state (D) in Fig.~\ref{fig:fig3}c. In this state, we have measured the pressure $P_D$ inside the droplet using different methods. First, we used direct computation from the diagonal components of the pressure tensor in a large portion of the liquid droplet. We confirmed that the potential energy and structural relaxation time correspond to the pressure $P_D$. We also measured the expansion of the droplet by following the displacement field in the radial direction. The resulting density $\rho_D$ agrees with the measured pressure $P_D$. At a given $\Ta$, simulations were performed for three different droplet sizes $R_0 = 50$, 80, 100 which gave identical results. We also checked that our results do not depend on the specific initial configuration.

\textbf{Simulations of macroscopic fronts.} For large fronts we again start from the expanded glass state (G) but we do not need to replicate the system in both directions. We directly perform high-temperature simulations to transform a slit of width $2W_0$ into a liquid, as shown in Fig.~\ref{fig:fig3}a. This is followed by some dynamics at $\Ta$. We then perform $NPT$ simulations at pressure $P_i$ and temperature $\Ta$. At a given $\Ta$, simulations were performed for three different widths $W_0 = 50$, 80, 100.  We also checked that our results do not depend on the specific initial configuration.

\textbf{Bond-breaking correlation function and phase definition.} We distinguish between the glass and liquid states using the bond-breaking correlation function given by
\begin{equation}
    \CB^i(t) = \frac{n_i \qty(t \vert 0)}{n_i(0)},
\end{equation}
where $n_i(t)$ is the number of neighbours of particle $i$ at time $t$, and
$n_i \qty(t \vert 0)$ is the number of neighbors of particle $i$ at time which were also neighbor at $t=0$. At $t=0$, the neighbors of particle $i$ are defined as all particles $j \neq i$ whose interparticle distance is smaller than a threshold, $r_{ij} / \sigma_{ij}<1.35$, which corresponds to the first minima of the rescaled radial distribution function $g(r/\sigma_{ij})$. We define liquid particles as those with $\CB^i(t) \leq 0.5$. The bond-breaking correlation function is defined from the ensemble average:
\begin{equation}
    \CB(t) = \expval{ \frac{1}{N} \sum_{i=1}^N \CB^i(t) },
\end{equation}
where $N$ is the number of particles. One can define from the bond-breaking correlation a characteristic time for the structural relaxation of the equilibrium bulk system as $\CB(t=\tau_B) = 0.5$.

\textbf{Self-intermediate scattering function.} We use another common correlation function to characterize the bulk dynamics, namely the self-intermediate scattering function, given by 
\begin{equation}
    F_s(t) = \expval{\frac{1}{N} \sum_{i=1}^N \cos[\vb{q}\cdot \delta \vb{r}_i(t)]},
\end{equation}
where $\delta \vb{r}_i = \vb{r}_i(t) - \vb{r}_i(0)$ and wave vectors with $\abs{\vb{q}}=6.9$, corresponding to the position of the global maximum in the static structure factor $S(\vb{q})$. We define the $\alpha$-relaxation time as $F_s(t=\tau_\alpha) = 1/e$. For large $2d$ systems $F_s(t)$ can be affected by Mermin-Wagner fluctuations, which depend logarithmically on the system size~\cite{Illing2017}. To avoid this effect, we performed bulk MD equilibrium simulations of a relatively small system with $N=2000$.

\textbf{Chord length distribution and characteristic length scale.} We employ the chord length distribution~\cite{Testard2011,Testard2014} to characterize the size of the liquid domains. Briefly, we discretize the spatial coordinates on a grids with cells of linear size $1.47$, so that the cell is large enough to ensure that there is at least one particle in the cell~\cite{scalliet2022thirty}. Then, we average $\CB^i$ over all the particles in a given cell. We binarize the result so that the cell is either in state 0 (liquid) when $\CB^i \le 0.5$ or 1 (glass) otherwise. The chord length $\ell$ is then defined from the intersection of segments in the $x$ and $y$ directions with the liquid domains. By averaging over all segements in both directions, we get the chord length distribution $P(\ell)$ and we extract a characteristic length scale as the first moment of this distribution. 

\clearpage

\newpage

\section*{Supplementary Information}

The transformation of a glass into a liquid has been previously described using a classical nucleation-and-growth Avrami dynamics~\cite{Jack2016}, which leads to the Avrami equation~\cite{Avrami1939}:
\begin{equation}
    \xl(t) = 1 - \exp(-K t^n),
\end{equation}
where $\xl$ is the liquid fraction at time $t$, $K$ and $n$ are parameters dependent on the transformation characteristics.

A more general expression for the phase transformation is obtained as follows. The number of nuclei $\dd N$ that appear during a time interval $\dd \tau$ is given by $\dd N = I V \dd \tau$, with $I$ the nucleation rate per unit volume, and $V$ is the volume of the system. One can suppose that the volume of each nucleus grows isotropically in time, with a radius $r(t)$. Then, the extended volume of the new phase resulting from the nucleation and growth of the nuclei that appeared in $\dd \tau$ is $\dd V_\mathrm{ext} = g_d r^d \dd N$, where $d$ is the system's dimension and $g_d$ a geometrical factor (\emph{e.g.} $g_d = 4\pi/3$ in 3$d$ and $\pi$ in 2$d$). This extended transformed volume defines the extended fraction $\dd x_\mathrm{ext} = \dd V_\mathrm{ext}/V$. However, part of this extended fraction covers already transformed material. Therefore, the actual liquid fraction that forms during an increment $\dd \tau$ will be proportional to the fraction of the untransformed phase: $\dd \xl = \dd x_\mathrm{ext} (1-\xl).$ Integrating this equation one obtains
\begin{equation}
    \xl = 1 - \exp(-x_\mathrm{ext}).
    \label{eq:Avrami_gal}
\end{equation}
The Avrami equation is a particular solution of Eq.~(\ref{eq:Avrami_gal}) in the case where the growth velocity $v$ and the nucleation rate $I$ are constant. From these assumptions, one can show that, for constant $I>0$ then $n=d+1$ (with $d$ the spatial dimension), while for $I=0$ and all nucleation sites are present at $t=0$ one gets $n=d$. Specifically, in the case $d=2$, a constant $I$ implies that circular domains grow with a linear size $r(t) \sim v t$, so that
\[ x_\mathrm{ext} = \int_0^t \pi I r^2(\tau) \dd \tau = \frac{\pi}{3} I v^2 t^3 ,\]
yielding the Avrami exponent $n=d+1=3$.

To understand the anomalous Avrami exponent ($n \approx 4.5$) obtained in our simulations, we introduce a simple modification to the above picture using insights from our MD simulations where we observe a non-constant growth velocity, which goes from $v_1$ observed for isolated droplets, to $v_2 > v_1$ observed for large fronts. We introduce the dimensionless parameter 
$\alpha = v_2 / v_1>1$ to quantify this effect.

\begin{figure}
\centering
\includegraphics[width=0.71\linewidth]{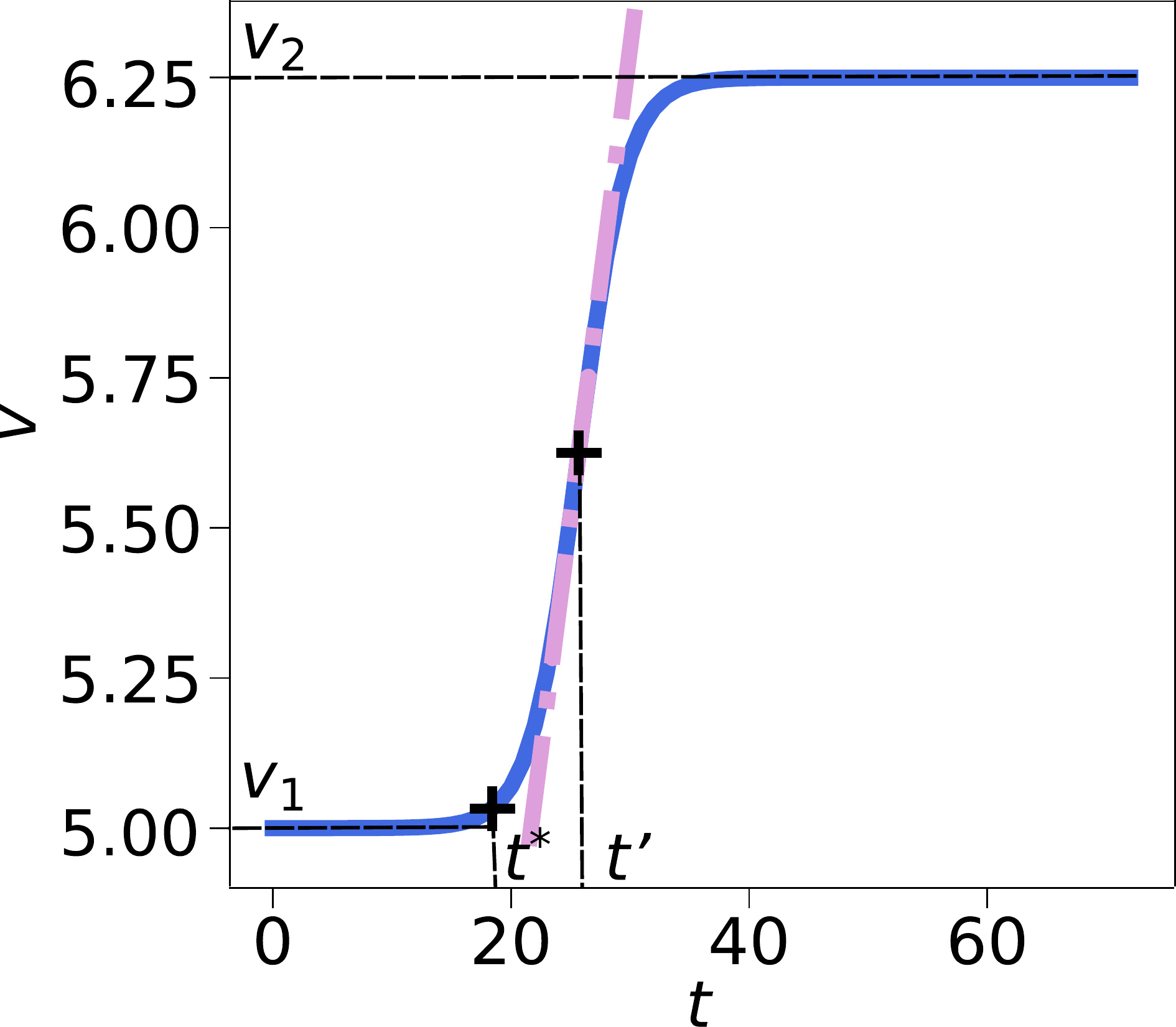}
\caption{{\bf Minimal model for time-dependent velocity.} We introduce a simple model where the velocity smoothly evolves from $v_1$ to $v_2>v_1$ at a typical timescale $t^*$, using a sigmoidal function, Eq.~(\ref{eq:vt_gal}), which is well approximated by a linear time dependence near the inflection point. Here, $v_1=5$, $v_2=6.25$, $I=8\times 10^{-7}$ and $\Delta t= 2$.}
\label{fig:figS1}
\end{figure}

We model the evolution of the growth velocity so that it smoothly evolves from $v_1$ at short times to $v_2$ at long times. To do so with minimal ingredients, we introduce a crossover length scale $R^*$, or, equivalently a crossover time scale $t^* = R^* / v_1$ controlling the velocity evolution. We estimate $R^*$ as the average distance between nucleated droplets~\cite{Jack2016}, namely
\[ R^* = \frac{ (v_1/I)^{1/(d+1)}}{2}, \]
so that $R^*$ is related to the typical droplet size when they start to merge to form large fronts. We also introduce a short duration $\Delta t$ for the crossover to occur leading to a sigmoidal functional form,   
\begin{equation}
    v(t) = v_1 \frac{1 + \alpha \exp(\frac{t - t'}{\Delta t})}{1 + \exp(\frac{t - t'}{\Delta t})};
    \label{eq:vt_gal}
\end{equation}
where $t'=t^* + 2 \Delta t$, and $t$ is the simulation time.
A specific example is shown in Fig.~\ref{fig:figS1}. 

In the intermediate time regime, when $t\sim t'$, one can perform the Taylor series of Eq.~(\ref{eq:vt_gal}) to get 
\[
    v(t\sim t') = v_1 \qty[ \frac{\alpha+1}{2} + \frac{1}{4}(\alpha-1) \frac{t - t'}{\Delta t} + \mathcal{O}(t^3)],
\]
allowing us to linearise the velocity in the transition regime as $v(t) \sim v_1 (at + b)$ with $a=\frac{\alpha - 1}{4\Delta t}$ and $b=\frac{\alpha+1}{2} - a t'$. In this regime, one gets 
\begin{equation}
\begin{split}
    x_\mathrm{ext} &= \int_0^t \pi I r^2(\tau) \dd \tau \\
    &= \pi I v_1^2 \qty[a^2 \frac{t^5}{5}+2 ab \frac{t^4}{4} + b^2 \frac{t^3}{3}],
\end{split}
\end{equation}
suggesting that an effective exponent larger than $3$ and close to $n \approx 5$ holds. 

\begin{figure}
\centering
\includegraphics[width=\linewidth]{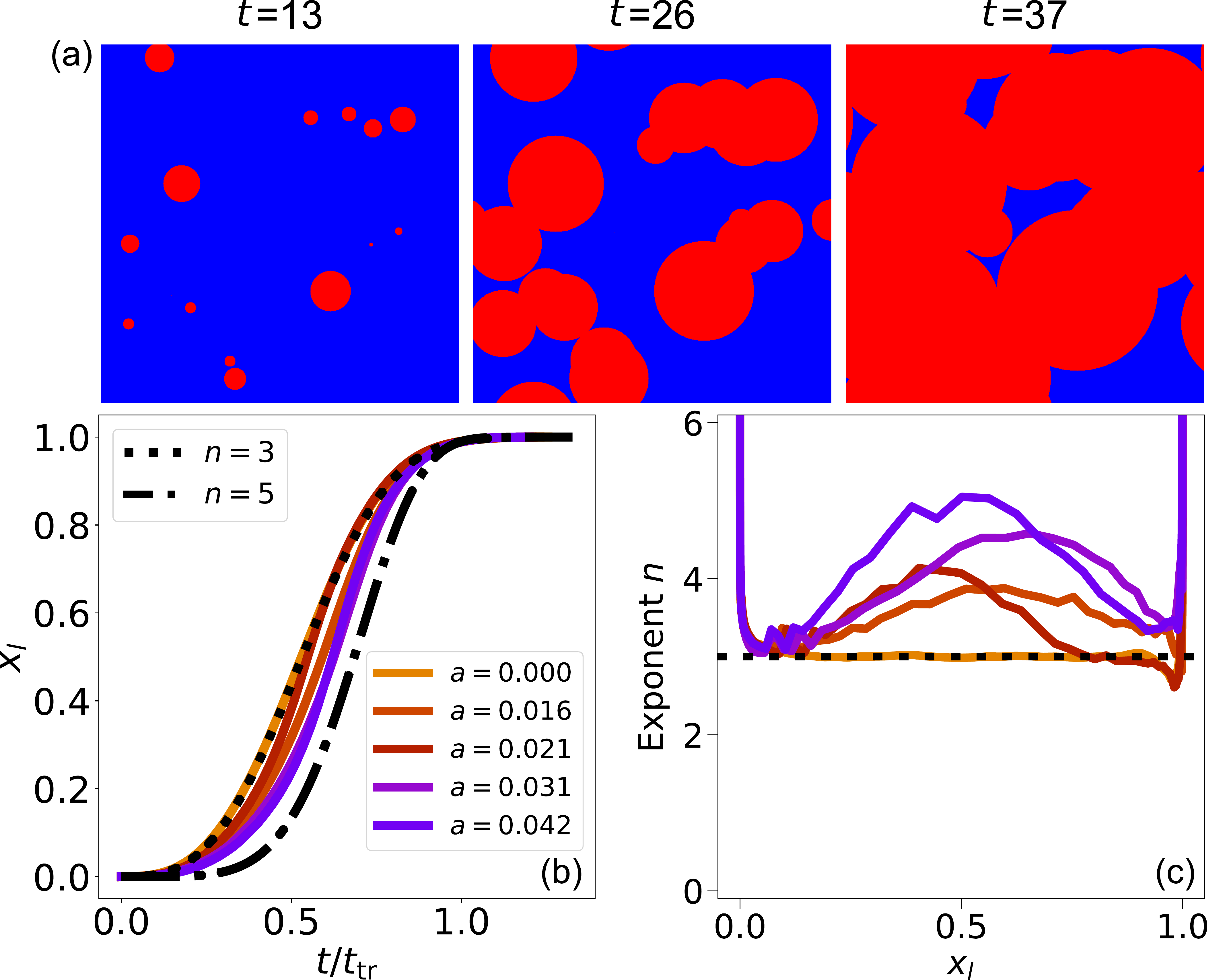}
\caption{{\bf Monte Carlo simulations.} (a) Snapshots of Monte Carlo simulations of the nucleation and growth dynamics with time-dependent velocity. The color codes for glass (blue) and liquid (red). The velocity time evolution is the same as in Fig.~\ref{fig:figS1}, yielding $a\sim 0.031$.
  (b) Liquid fraction versus normalized time for different values of $a$, which characterizes the steepness of the velocity change, along with different Avrami functions.
  (c) Effective Avrami exponent $n$ from the data in (b).}
\label{fig:figS2}
\end{figure}

To confirm these analytical results, we have performed Monte Carlo simulations of this kinetics, as illustrated in Fig.~\ref{fig:figS2}a. In these simulations,
we introduce a discrete lattice. The liquid nucleates stochastically with a constant nucleation rate $I$ on the sites of the lattice. The nucleated sites then grow circular domains with a velocity given by Eq.~(\ref{eq:vt_gal}), independently of other liquid domains. The model has four parameters: the nucleation rate $I$, the two velocities $v_1$ and $v_2 = \alpha v_1$ and the timescale $\Delta t$ controlling the sharpness of the change between these two values. We find, however, that it is the parameter $a = (\alpha-1)/\Delta t$, which characterizes the steepness of the velocity change, which mostly controls the obtained kinetics. 

We can extract the time evolution of the fraction of liquid $x_l(t)$ and the effective Avrami exponent $n$ from these Monte Carlo simulations for different values of $a$, as shown in Fig.~\ref{fig:figS2}b. For $a=0$, the velocity is constant throughout the simulation, we recover the classical Avrami kinetics, yielding $n=3$, see Fig.~\ref{fig:figS2}c. However, as soon as the velocity changes, one finds a larger Avrami exponent, going up to $n=5$ for $a=0.042$.

\begin{figure}
\centering
\includegraphics[width=\linewidth]{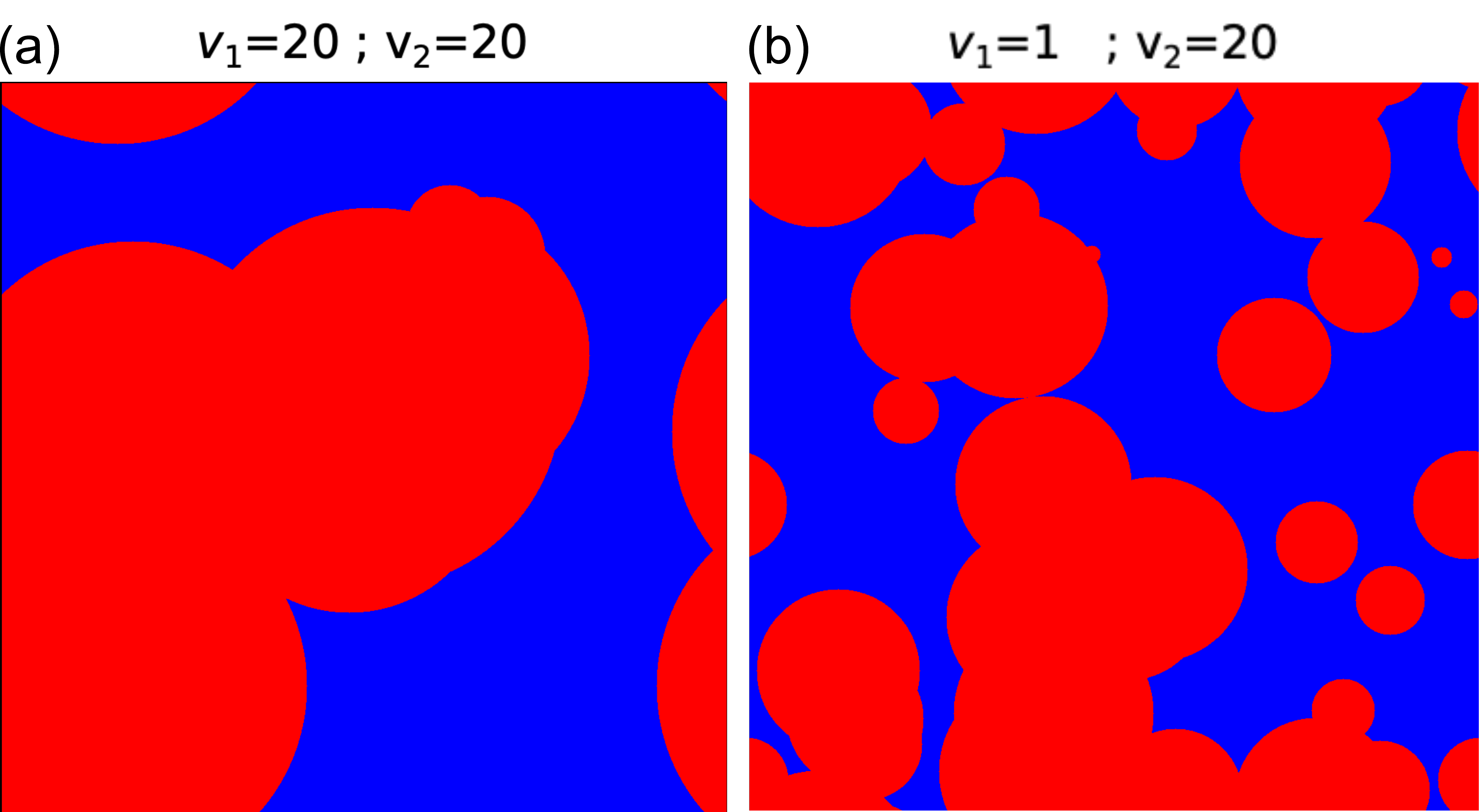}
\caption{{\bf Characteristic length scale in Monte Carlo simulations.} Snapshots of two different Monte Carlo simulations for $I=5\times 10^{-7}$ and $\Delta t = 5$, taken at $t=t_{1/2}$. The characteristic distance between nucleation events, and thus $\lphtr$, is smaller when $v_1 < v_2$ (b) compared to the standard Avrami case where $v_1=v_2$ (a). }
\label{fig:figS3}
\end{figure}

Finally, it is interesting to use this minimal model to explore the two characteristic lengths discussed in the main text, $\lphtr$ and $\lcross$. When $v_1=v_2$, we expect that $\lphtr = \lcross$. However, by decreasing the value of $v_1$ to explore the regime $v_1 < v_2$ (keeping all other parameters unchanged), the transformation time becomes longer and thus $\lcross = v_2 \ttransf$ increases. Nevertheless, as discussed in the main text, larger transformation times at constant nucleation rate implies the presence of a larger number of nuclei at time $t_{1/2}$, decreasing the distance between them and therefore decreasing $\lphtr$. The generic expectation that $\lphtr<\lcross$ is consistent with our MC simulations. In Fig.~\ref{fig:figS3} we show two different simulations with the same nucleation frequency but distinct velocity evolution. One can directly observe how decreasing $v_1$ results in a significantly smaller characteristic length $\lphtr$.

This minimal analytic model where a single modification to the Avrami picture is introduced is therefore able to reproduce the results obtained from the MD simulations described in the main text.

\clearpage

\end{document}